\documentclass{birkjour}
\usepackage{amsthm}
  \usepackage{amssymb}
  \usepackage{amsmath}
  \usepackage{amscd}
  \usepackage{eucal}
  \usepackage{mathrsfs}
  \usepackage{upgreek}
   \usepackage{dsfont}
  \usepackage[T1]{fontenc}
  \usepackage{aurical}
  \usepackage{bbm}
\usepackage{array}

\newcommand{\E}{\mathfrak{E}}

\newcommand{\D}{\mathfrak{D}}

\newcommand{\A}{\mathfrak{A}}

\newcommand{\BV}{\mathfrak{BV}}

\newcommand{\1}{\mathds{1}}

\newcommand{\dc}{\mathcal{O}} 

\newcommand{\La}{\Lambda}

\newcommand{\Obj}{\mathrm{Obj}}

\newcommand{\Nat}{\mathrm{Nat}}

\newcommand{\Loc}{\mathrm{\mathbf{Loc}}} 
\newcommand{\Obs}{\mathrm{\mathbf{Obs}}} 
\newcommand{\Vect}{\mathrm{\mathbf{Vec}}} 

\newcommand{\be}{\begin{equation}}
\newcommand{\ee}{\end{equation}}

\newcommand{\RR}{\mathbb{R}} 
\newcommand{\CC}{\mathbb{C}} 

\newcommand{\Lcal}{\mathcal {L}}

\newcommand{\Ncal}{\mathcal{N}}
\newcommand{\Mcal}{\mathcal{M}}

 \theoremstyle{definition}
 
 \theoremstyle{remark}

 \numberwithin{equation}{section}

\begin{document}

%
%
%
%
%
%
%
%
%

\title{Local covariance and background independence}
\author
{Klaus Fredenhagen}

\address{ II Inst. f. Theoretische Physik,\\
Universit\"at Hamburg,\\
Luruper Chaussee 149, \\
D-22761 Hamburg, Germany}

\email{klaus.fredenhagen@desy.de}

\author{Katarzyna Rejzner}
\address{ II Inst. f. Theoretische Physik,\\
Universit\"at Hamburg,\\
Luruper Chaussee 149, \\
D-22761 Hamburg, Germany}
\email{katarzyna.rejzner@desy.de}
\subjclass{}

\keywords{}

\date{September 28, 2010}

\begin{abstract}
One of the many conceptual difficulties in the development of quantum gravity is the role of a background geometry for the structure of quantum field theory. To some extent the problem can be solved by the principle of local covariance. The principle of local covariance was originally imposed in order to restrict the renormalization freedom for quantum field theories on generic spacetimes. It turned out that it can also be used to implement the request of background independence. Locally covariant fields then arise as background independent entities.
\end{abstract}

\maketitle
\section{Introduction}
The formulation of a theory of quantum gravity is one of the most important unsolved problems in physics. 
It faces not only technical but above all, conceptual problems. The main one arises from the fact that, 
in quantum physics, space and time are a priori structures which enter the definition of the theory as well as its interpretation in a crucial way. On the other hand, in general relativity, spacetime is a dynamical object, determined by classical observables. To solve this apparent discrepancy, radical new approaches were developed. Among these  the best known are string theory and loop quantum gravity. Up to now all these approaches meet the same problem: It is extremely difficult to establish the connection to actual physics. 

Instead of following the standard approaches to quantum gravity we propose a more conservative one.  We concentrate on the situation when the influence of the gravitational field is weak. This idealization is justified in a large scope of physical situations. Under this assumption one can approach the problem of quantum gravity from the field-theoretic side. In the first step we consider spacetime to be a given Lorentzian manifold, on which quantum fields live. In the second step gravitation is quantized around a given background. This is where the technical problems start. The arising theory is nonrenormalizable, in the sense that infinitely many counter terms arise in the process of renormalization. Furthermore, the causal structure of the theory is determined by the background metric. Before discussing these difficulties we want to point out that also
the first step is by no means trivial. Namely, the standard formalism of quantum field theory is based on the symmetries of Minkowski space. Its generalization even to the most symmetric spacetimes (de Sitter, anti-de Sitter) poses problems. There is no vacuum, no particles no S-matrix, etc.\ldots Solution to these difficulties is provided by concepts of algebraic quantum field theory and methods from microlocal analysis.

One starts with generalizing the Haag-Kastler axioms to generic spacetimes. We consider
algebras $\A(\dc)$ of observables which can be measured within the spacetime region $\dc$, satisfying the axioms of isotony, locality (commutativity at spacelike distances) and covariance. Stability is formulated as the existence of a vacuum state (spectrum condition). The existence of a dynamical law (field equation) is understood as fulfilling the timeslice axiom (primitive causality) which says that the algebra of a timeslice is already the algebra of the full spacetime. This algebraic framework, when applied to generic Lorentzian manifolds, still meets a difficulty. The causal structure is well defined but the absence of nontrivial symmetries rises the question: What is the meaning of repeating an experiment? This is a crucial point if one wants to keep  the probability interpretation of quantum theory. A related issue is the need of a generally covariant version of the spectrum condition. These problems can be solved within \textit{locally covariant quantum field theory}, a new framework for QFT on generic spacetime proposed in \cite{BFV}.
\section{Locally covariant quantum field theory}
The framework of locally covariant quantum field theory was developed in \cite{BFV,HW1,HW2}. The idea is to construct the theory simultaneously on all spacetimes (of a given class) in a coherent way.
Let $\Mcal$ be a  globally hyperbolic, oriented, time oriented Lorentzian 4d spacetime.
Global hyperbolicity means that $\Mcal$ is diffeomorphic to $\RR \times \Sigma$, where $\Sigma$ is a Cauchy surface of $\Mcal$. Between spacetimes one considers a class of admissible embeddings.
An embedding $\chi:\Ncal\rightarrow  \Mcal$ is called admissible, if it is isometric, time orientation and orientation preserving, and causally convex in the following sense: If $\gamma$ is a causal curve in $\Mcal$ with endpoints $p,q\in\chi(\Ncal)$ then $\gamma=\chi\circ\gamma'$ with a causal curve $\gamma'$ in $\Ncal$.
A locally covariant QFT is defined by assigning to spacetimes  $\Mcal$ corresponding unital $C^*$-algebras $\A(\Mcal)$. This assignment has to fulfill a set of axioms, which generalize the Haag-Kastler axioms:
\begin{enumerate}
\item $\Mcal \mapsto \A(\Mcal)$ unital $C^*$-algebra (\textit{local observables}),
\item	If $\chi: \Ncal \rightarrow \Mcal$ is an admissible embedding, then $\alpha_\chi:\ \A(\Ncal)\rightarrow \A(\Mcal)$ is a unit preserving $C^*$-homomorphism (\textit{subsystems}),
\item	Let $\chi:\Ncal \rightarrow \Mcal$, $\chi':\Mcal\rightarrow \Lcal$ be admissible embeddings, then
$\alpha_{\chi'\circ\chi} = \alpha_{\chi'}\circ\alpha_\chi$ (\textit{covariance}),
\item If $\chi_1 : \Ncal_1 \rightarrow \Mcal$, $\chi_2 : \Ncal_2 \rightarrow \Mcal$ are admissible embeddings such that $\chi_1(\Ncal_1)$ and $\chi_2(\Ncal_2)$ are spacelike separated in $\Mcal$ then
$[\alpha_{\chi_1} (\A(\Ncal_1)), \alpha_{\chi_2} (\A(\Ncal_2))] = {0}$ (\textit{locality}),
\item	If $\chi(\Ncal)$ contains a Cauchy surface of $\Mcal$ then $\alpha_\chi(\A(\Ncal )) = \A(\Mcal)$ (\textit{timeslice axiom}).
\end{enumerate}
Axioms 1-3 have a natural interpretation in the language of category theory. Let $\Loc$ be the category of globally hyperbolic Lorentzian spacetimes with admissible embeddings as morphisms and $\Obs$ the category of unital $C^*$-algebras with homomorphisms as morphisms. Then a locally covariant quantum field theory is defined as a covariant functor $\A$ between $\Loc$ and $\Obs$, with $\A\chi:=\alpha_\chi$. 

The fourth axiom is related to the tensorial structure of the underlying categories. The one for the category $\Loc$ is given in terms of disjoint unions. It means that objects in $\Loc^{\otimes}$ are all elements $\Mcal$ that can be written as $\Mcal_1\otimes\ldots\otimes\Mcal_N:=\Mcal_1\coprod \ldots\coprod \Mcal_n$ with the unit provided by the empty set $\varnothing$. The admissible embeddings are maps $\chi: \Mcal_1\coprod \ldots\coprod \Mcal_n\rightarrow  \Mcal$ such that each component satisfies the requirements mantioned above  and additionally all images are spacelike to each other, i.e., $\chi(\Mcal_1) \perp\ldots\perp(\Mcal_n)$. The tensorial structure of the category $\Obs$ is a more subtle issue. Since there is no unique tensor structure on general locally convex vector spaces, one has to either restrict to some subcategory of $\Obs$ (for example nuclear spaces) or make a  choice of the tensor structure basing on some physical requirements. The functor $\A$ can be then extended to a functor $\A^\otimes$ between the categories $\Loc^\otimes$ and $\Obs^\otimes$. It is a covariant tensor functor if it holds:
\begin{eqnarray}
\A^\otimes\left(\Mcal_1\coprod\Mcal_2\right)&=&\A(\Mcal_1)\otimes\A(\Mcal_2)\\
\A^\otimes(\chi\otimes\chi')&=&\A^\otimes(\chi)\otimes\A^\otimes(\chi')\\
\A^\otimes(\varnothing)&=&\CC
\end{eqnarray}
It can be shown, that if $\A$ is a tensor functor, then the causality follows. To see this consider the natural embeddings $\iota_{i}:\Mcal_i\rightarrow\Mcal_1\coprod\Mcal_2$, $i=1,2$ for which $\A\iota_1(A_1)=A_1\otimes\1$, $\A\iota_2(A_2)=\1\otimes A_2$, $A_i\in\A(\Mcal_i)$. Now let $\chi_i:\Mcal_i\rightarrow\Mcal$ be admissible embeddings such that the images of  $\chi_1$ and  $\chi_2$ are causally disjoint in $\Mcal$. We define now an admissible embedding $\chi:\Mcal_1\coprod\Mcal_2\rightarrow\Mcal$ as:
\be
\chi(x)=\left\{
\begin{array}{lcl}
\chi_1(x)&,&x\in\Mcal_1\\
\chi_2(x)&,&x\in\Mcal_2
\end{array}
\right.
\ee
Since $\A^\otimes$ is a covariant tensor functor, it follows:
\be
[\A\chi_1(A_1),\A\chi_2(A_2)]=\A\chi[\A\iota_1(A_1),\A\iota_2(A_2)]=\A\chi[A_1\otimes\1,\1\otimes A_2]=0
\ee
This proves the causality. With a little bit more work it can be shown that also the opposite implication holds, i.e. the causality axiom implies that the functor $\A$ is tensorial.

The last axiom is related to cobordisms of Lorentzian manifolds. One can associate to a Cauchy surface $\Sigma\subset\Mcal$ a family of algebras $\{\A(\Ncal)\}_{\Ncal\in I}$, where the index set consists of all admissibly embedded subspacetimes $\Ncal$ of 
$\Mcal$, that contain the Cauchy surface $\Sigma$. On this family we can introduce an order relation $\geq$, provided by the inclusion. Let $\Ncal_i,\Ncal_j\in I$, such that $\Ncal_i\subset\Ncal_j\in I$, then we say that $\Ncal_i\geq\Ncal_j$. Clearly $\Sigma$ is the upper limit with respect to the order relation $\geq$, hence we obtain a directed system of algebras  $(\{\A(\Ncal)\}_{\Ncal\in I},\geq)$.
Now let $\chi_{ij}:\Ncal_i\hookrightarrow\Ncal_j$ be the canonical isometric embedding of $\Ncal_i\geq\Ncal_j$. From the covariance it follows that there exists a morphism of algebras $\alpha_{\chi_{ji}}:\A(\Ncal_i)\hookrightarrow\A(\Ncal_j)$. We can now consider a family of all such mappings between the elements of the directed system  $(\{\A(\Ncal)\}_{\Ncal\in I},\geq)$. Clearly $\alpha_{\chi_{ii}}$ is the identity on $\A(\Ncal_i)$ and $\alpha_{\chi_{ik}} = \alpha_{\chi_{ij}}\circ\alpha_{\chi_{jk}}$ for all $\Ncal_i\leq\Ncal_j\leq\Ncal_k$. This means that the family of mappings $\alpha_{\chi_{ij}}$ provides the transition morphisms for the directed system  $(\{\A(\Ncal)\}_{\Ncal\in I},\geq)$ and we can define the projective (inverse) limit of the inverse system of algebras  $(\{\A(\Ncal)\}_{\Ncal\in I},\geq,\{\alpha_{\chi_{ij}}\})$, i.e.:
\be
\A(\Sigma):=\!\varprojlim\limits_{\Ncal\supset\Sigma}\A(\Ncal)= \Big\{\mbox{ germ of }(a)_I\!\! \in\!\! \prod_{{\Ncal\in I}}\A(\Ncal) \;\Big|\; a_{\Ncal_i}= \alpha_{\chi_{ij}}( a_{\Ncal_j})\ \forall\ \Ncal_i \leq \Ncal_j \Big\}. 
\ee
The algebra $\A(\Sigma)$ obtained in this way depends in general on the germ of $\Sigma$ in $\Mcal$.
If we consider natural embeddings of Cauchy surfaces $\Sigma$ in  $\Mcal$, then, acting with the functor $\A$ we obtain homomorphisms of algebras, which we denote by $\alpha_{\Mcal\Sigma}$. The time-slice axiom implies that these homomorphisms are in fact isomorphisms. It follows that the propagation from $\Sigma_1$ to another Cauchy surface $\Sigma_2$ is described by the isomorphism:
\be\label{alphaM}
\alpha^{\Mcal}_{\Sigma_1\Sigma_2}:=\alpha_{\Mcal\Sigma_1}^{-1}\alpha_{\Mcal\Sigma_2} \ .
\ee
Givan a cobordism, i.e. a Lorentzian manifold $\Mcal$ with future/past boundary $\Sigma_\pm$ we obtain an assignment: $\Sigma_\pm\mapsto\A(\Sigma_\pm)$, $\Mcal\mapsto\alpha^{\Mcal}_{\Sigma_-\Sigma_+}$.
The concept of relative Cauchy evolution obtained in this way realizes the notion of dynamics in the locally covariant quantum field theory framework. This provides a solution to the old problem of Schwinger to formulate the functional evolution of the quantum state. The original idea to understand it as a unitary map between Hilbert spaces turned out not to be a viable concept even in Minkowski spacetime \cite{Torr}. Nevertheless one can understand the dynamical evolution on the algebraic level as an isomorphism of algebras corresponding to the Cauchy surfaces. This idea was already applied by Buchholz and Verch \cite{Ver} to some concrete examples and the locally covariant quantum theory provides a more general framework in which this approach is justified. Note also the structural similarity to topological field theory \cite{Seg}. There, however, the objects are finite dimensional vector spaces, so the functional analytic obstructions which are typical for quantum field theory do not arise.
\section{Perturbative quantum gravity}
After the brief introduction to the locally covariant QFT framework we can now turn back to the problem of quantum gravity seen from the point of view of perturbation theory. First we split of the metric:
\be
g_{ab} = g^{(0)}_{ab} + h_{ab}\,,
\ee
where $g^{(0)}$ is the background metric, and $h$ is a quantum field. Now we can
renormalize the Einstein-Hilbert action by the Epstein-Glaser
method (interaction restricted to a compact region between two
Cauchy surfaces) and construct the functor $\A$. Next we compute (\ref{alphaM}) for two background metrics which differ by $\kappa_{ab}$ compactly supported between two Cauchy surfaces. Let $\Mcal_1=(M, g^{(0)})$ and  $\Mcal_2=(M, g^{(0)}+\kappa)$. Following \cite{BFV,BF1} we assume that there are two causally convex neighbourhoods $\Ncal_\pm$ of  the Cauchy surfaces $\Sigma_\pm$, which can be admissibly embedded both in $\Mcal_1$ and $\Mcal_2$ and $\kappa$ is supported in a compact region between $\Ncal_-$ and $\Ncal_+$. We denote the corresponding embeddings by $\chi_{i}^\pm:\Ncal_\pm\rightarrow\Mcal_i$, $i=1,2$. We can now define an automorphism of $\Mcal_1$ by:
 \be
 \beta_\kappa:=\alpha_{\chi^-_1}\circ\alpha_{\chi^-_2}^{-1}\circ\alpha_{\chi^+_2}\circ\alpha_{\chi^+_1}^{-1}\,.
\ee
This automorphism corresponds to a change of the background between the two Cauchy surfaces. Under the geometrical assumptions given in \cite{BFV} one can calculate a functional derivative of $\beta_\kappa$ with respect to $\kappa$. If the metric is not quantized it was shown in \cite{BFV} that this derivative corresponds to the commutator with the stress-energy tensor. In case of quantum gravity $\frac{\delta  \beta_\kappa}{\delta\kappa_{ab}(x)}$
 involves in addition also the Einstein tensor. Therefore the background independence may be formulated as the condition that $\frac{\delta  \beta_\kappa}{\delta\kappa_{ab}(x)}=0$, i.e. one requires the validity of Einstein's equation for the quantized fields. This can be translated into a corresponding renormalization condition.

The scheme proposed above meets some technical obstructions. First of them is the nonrenormalizability. This means that in every order new counter terms appear. Nevertheless, if these terms are sufficiently small, we can still have a predictive power of the resulting theory, as an effective theory. The next technical difficulty is imposing of the constraints related to the gauge (in this case diffeomorphism) invariance. In perturbation theory this can be done using the BRST method \cite{BRST1,BRST2} (or more generally Batalin-Vilkovisky formalism \cite{Batalin:1981jr,HennBar}). Since this framework is based on the concept of local objects, one encounters another problem.
Local BRST cohomology turns out to be trivial \cite{FR}, hence one has to generalize the existing methods to global objects. Candidates for global quantities are \textit{fields}, considered as natural transformations between the functor of test function spaces $\D$ and the quantum field theory functor $\A$. A quantum field $\Phi:\D\rightarrow \A$ corresponds therefore to a family of mappings $(\Phi_\Mcal)_{\Mcal\in\Obj(\Loc)}$, such that $\Phi_\Mcal(f)\in\A(\Mcal)$ for $f\in\D(M)$ and given a morphism $\chi: \Ncal \rightarrow \Mcal$ we have $\alpha_\chi(\Phi_\Ncal(f)) = \Phi_\Mcal(\chi_*f)$.
\section{BRST cohomology for classical gravity}
While quantum gravity is still elusive, classical gravity is (to some extent) well understood.
Therefore one can try to test concepts for quantum gravity in a suitable framework for classical gravity.
Such a formalism is provided by the algebraic formulation, where classical field theory occurs as the $\hbar= 0$ limit of quantum field theory \cite{BDF,DF,DF02,DF04}. In this approach (the \textit{functional approach}) one replaces the associative involutive algebras by Poisson algebras. In case of gravity, to obtain a suitable Poisson structure one has to fix the gauge. In the BRST method this is done by adding a gauge fixing term and a ghost term to the Einstein-Hilbert action. The so called \textit{ghost fields} have a geometrical interpretation as
Maurer-Cartan forms on the diffeomorphism group. This can be made precise in the framework of infinite dimensional differential geometry. The notion of infinite dimensional manifolds and in particular, infinite dimensional Lie groups is known in mathematics since the reviews of Hamilton \cite{Ham} and Milnor \cite{Mil}. Because one needs to consider manifolds modeled on general locally convex vector spaces, an appropriate calculus has to be chosen. Unfortunately the choice is not unique when we go beyond the Banach spaces. Historically the earliest works concerning such generalization of calculus are those of Michal \cite{Mich} (1938) and Bastiani \cite{Bast} (1964). At present there are two main frameworks in which the problems of infinite dimensional differential geometry can be approached: the convenient setting of global analysis \cite{FK,Michor} and the locally convex calculus \cite{Ham,Neeb}. Up to now both calculi coincide in the examples which were considered. 

First we sketch the BRST construction performed on the fixed background $\Mcal$. The basic objects of the classical theory are:
\begin{itemize}
\item $S$, a diffeomorphism invariant action,
\item Field content: configuration space $\E(\Mcal)$, considered as an infinite dimensional manifold: scalar, vector, tensor and spinor fields (including the metric), gauge fields,
\item Ghost fields (fermions): forms on the gauge algebra $\Gamma TM$, i.e. elements of $(\Gamma TM)^*$,
\item Antifields (fermions): vector fields $\Gamma T\E(\Mcal)$ on the configuration space,
\item Antifields of ghosts (bosons): compactly supported vector fields $\Gamma_c TM$.
\end{itemize}
The fields listed above constitute the minimal sector of the theory.
To impose a concrete gauge, one can also introduce further fields, the so called nonminimal sector.  For the harmonic gauge it consists of Nakanishi-Lautrup fields (bosonic) and antighosts (fermionic). The minimal sector of the BRST-extended functional algebra takes the form:
\be
\BV(\Mcal)=Sym( \Gamma_c T\Mcal)\widehat{\otimes}\Lambda(\Gamma T\E(\Mcal))\widehat{\otimes}\La(\Gamma T\Mcal)^*\,,\label{BVfix}
\ee
where $\widehat{\otimes}$ denotes the sequentially completed tensor product, and $Sym$ is the symmetric algebra. Algebra (\ref{BVfix}) is equipped with a grading called \textit{the ghost number} and a graded differential $s$ consisting of two terms $s=\delta+\gamma$. Both $\delta$ and $\gamma$ are graded differentials.
The natural action of $\Gamma T\Mcal$ by the Lie derivative on $\E(\Mcal)$ induces in a natural way an action  on $T\E(\Mcal)$. Together with the adjoint action on $\Gamma_c T\Mcal$ we obtain an action of $\Gamma T\Mcal$ on $Sym( \Gamma_c T\Mcal)\otimes\Lambda(\Gamma T\E(\Mcal))$ which we denote by $\rho$. We can now write down how  $\delta$ and $\gamma$ act on the basic fields $a\in \Gamma_c T\Mcal$, $Q\in\Gamma T\E(\Mcal)$, $\omega\in(\Gamma T\Mcal)^*$:
\begin{itemize}
\item $\langle\gamma(a\otimes Q\otimes\1),X\rangle:=\rho_X(a\otimes Q\otimes\1)$,
\item $\langle\gamma(a\otimes Q\otimes \omega),X\wedge Y\rangle:=\rho_X(a\otimes Q\otimes\langle \omega,Y\rangle)-\rho_Y(a\otimes Q\otimes\langle \omega,X\rangle)-a\otimes Q\otimes\langle \omega,[X,Y]\rangle$,
\item $\delta (\1\otimes Q\otimes \omega):=\1\otimes\partial_QS\otimes \omega$,
\item $\delta (a\otimes \1\otimes \omega):=\1\otimes\rho(a)\otimes \omega$.
\end{itemize}
Up to now the construction was done on the fixed spacetime, but it is not difficult to see that the assignment of the graded algebra $\BV(\Mcal)$ to spacetime $\Mcal$ can be made into a covariant functor \cite{FR}.

As indicated already, the BRST method when applied to gravity, has to be generalized to global objects. Otherwise the cohomology of the BRST operator $s$ turns out to be trivial. This corresponds to the well known fact that there are no local on-shell observables in general relativity. It was recently shown in \cite{FR} that one can introduce the BRST operator on the level of natural transformations and obtain in this way a nontrivial cohomology. Fields are now understood as natural transformations.  Let $\D^k$ be a functor from the category $\Loc$ to the product category ${\Vect}^k$, that assigns to a manifold $M$ a $k$-fold product of the test section spaces $\D(M)\times\ldots\times \D(M)$. Let $\Nat(\D^k,\BV)$ denote the set of natural transformations from $\D^k$ to $\BV$. We define the extended algebra of fields as:
\be
Fld=\bigoplus\limits_{k=0}^\infty \Nat(\D^k,\BV)\,,
\ee
It is equipped with a graded product defined as:
\be\label{ntprod}
(\Phi\Psi)_M(f_1,...,f_{p+q})=\frac{1}{p!q!}\sum\limits_{\pi\in P_{p+q}}\mathrm\Phi_M(f_{\pi(1)},...,f_{\pi(p)})\Psi_M(f_{\pi(p+1)},...,f_{\pi(p+q)})\,,
\ee
where the product on the right hand side is the product of the algebra $\BV(M)$. Let $\Phi$ be a field, then the action of the BRST differential on it is defined as:
\be
(s\Phi)_M(f):=s(\Phi_M(f))+(-1)^{|\Phi|}\Phi_M(\pounds_{(.)}f)\,,
\ee
where $|.|$ denotes the ghost number and the action of $s$ on $\BV(\Mcal)$ is given above. The physical fields are identified with the $0$-th cohomology of $s$ on $Fld$. Among them we have for example scalars constructed covariantly from the metric.
\section{Conclusions}
It was shown that a construction of quantum field theory on generic Lorentzian spacetime is possible, in accordance with the principle of general covariance. This framework can describe a wide range of physical situations. Also a consistent incorporation of the quantized gravitational field seems to be possible. Since the theory is invariant under the action of an infinite dimensional Lie group, the framework of infinite dimensional differential geometry plays an important role. It provides the mathematical setting in which the BV method has a clear geometrical interpretation. The construction of a locally covariant theory of gravity in the proposed setting was already performed for the classical theory. Basing on the gained insight it seems to be possible to apply this treatment also in the quantum case. One can then investigate the relations 
 to other field theoretical approaches to quantum gravity (Reuter\cite{Reut}, Bjerrum-Bohr \cite{BB},\ldots). As a conclusion we want to stress, that quantum field theory should be taken serious as a third way to quantum gravity.

\end{document}